\documentclass[prb,twocolumn,showpacs,preprintnumbers,amsmath,amssymb]{revtex4}
\usepackage{graphicx}
\usepackage{dcolumn}
\usepackage{bm}
\usepackage[tight]{subfigure}
\usepackage{enumerate}
\usepackage{amsmath}
\usepackage{verbatim}
\usepackage{color}

\begin{document}
\title{Using hybrid topological-spin qubit systems for two-qubit-spin gates}
\author{Martin Leijnse}
\author{Karsten Flensberg}

\affiliation{
  Center for Quantum Devices, Niels Bohr Institute,
  University of Copenhagen,
  2100~Copenhagen \O, Denmark 
}

\begin{abstract}
We investigate a hybrid quantum system involving spin qubits, based on the spins of electrons confined in quantum dots, 
and topological qubits, based on Majorana fermions. 
In such a system, gated control of the charge on the quantum dots allows transfer of quantum information between 
the spin and topological qubits, and the topological system can be used to facilitate transfer of spin qubits between 
spatially separated quantum dots and to initialize entangled spin-qubit pairs. 
Here, we show that the coupling to the topological system also makes it possible to perform entangling two-qubit gates 
on spatially separated spin qubits. 
The two-qubit gates are based on a combination of topologically protected braiding operations, 
gate-controlled charge transfer between the dots and edge Majorana modes, and measurements of the state 
of the topological qubits.
\end{abstract}
\maketitle
\section{Introduction.}
Spin qubits defined in quantum dots~\cite{Loss98} is a promising candidate for quantum information 
processing because of long coherence times and efficient implementation of one- and two-qubit gates,
as demonstrated in a number of recent experiments~\cite{Koppens05, Petta05, Koppens06, Nowack07}.
Single-qubit rotations can be implemented with time-dependent magnetic fields~\cite{Koppens06}, or with electric fields
utilizing the spin-orbit coupling~\cite{Nowack07}.
Two-qubit gates can be implemented by controlling the exchange interaction in a double dot~\cite{Loss98}, which is 
accomplished through gated control of the inter-dot tunnel coupling. The exchange interaction can also be used 
for single-qubit rotations, if the qubit is defined in some two-dimensional subspace of the two double 
dot spins~\cite{DiVincenzo00, Petta05}, e.g., the singlet and one of the triplets.
Readout can be done via spin-to-charge conversion, where one spin qubit state is prohibited from tunneling out of the dot,
either by a Zeeman splitting~\cite{Elzerman04} (in a single-spin qubit) or by Pauli spin-blockade~\cite{Petta05} 
(in a singlet-triplet qubit).

One problem with spin qubits is the limited range of the exchange interaction, causing problems with scaling 
to larger (many-qubit) systems~\cite{Svore05}. Similar problems are encountered in many other qubit realizations 
since it is hard to achieve long coherence times for a type of qubit which is sensitive to long-ranged interactions. 
Therefore, much theoretical and experimental work has been aimed at designing
quantum systems which can facilitate coupling between two (or more) spatially separated qubits, thus acting as 
a quantum bus for the primary qubits. An early example is the proposal to use phonons to couple qubits defined in 
trapped ions~\cite{Cirac95}. More recently, coupling of superconducting phase~\cite{Sillanpaa07} and charge~\cite{Majer07} 
qubits via a microwave field (circuit quantum electrodynamics) has been achieved experimentally. 
The same principles have been considered for spin qubits~\cite{Imamoglu99, Childress04, Jin12}, 
but this case is experimentally more problematic because of the rather weak coupling between the electron spin 
and the cavity field. Other theoretical suggestions involve e.g., coupling to extended spin chains~\cite{Friesen07}.

In this work we study the possibility of using a topological superconductor (TS) as a quantum bus for spin qubits. 
A TS is a superconductor with $p$-wave-type pairing, which has been argued to be realized e.g., in 
Sr$_2$RuO$_4$~\cite{DasSarma06}. Recently it was shown that $p$-wave type pairing may occur
in topological insulators ~\cite{Fu08} and strong spin-orbit 
semiconductors~\cite{Sau10, Alicea10, Oreg10, Lutchyn10} with induced magnetism and superconductivity. 
Perhaps the most spectacular aspect of TS is the Majorana fermion~\cite{Wilczek09} (MF) quasiparticles which form in vortices and
at edges, first discussed in the context of the $\nu = 5/2$ fractional quantum Hall state~\cite{Moore91}. 
Basically, a MF is "half of a Dirac fermion" and two MFs together form one fermionic degree of freedom, which may 
be highly nonlocal if the MFs are spatially separated. This non-locality offers protection from many types of decoherence. 
In topological quantum computation schemes~\cite{Kitaev03,Nayak08rev}, these non-local degrees of freedom are manipulated 
by braiding operations, i.e., by physical exchange of the associated 
\emph{local} quasiparticle excitations. This is possible because MFs satisfy non-abelian (non-commutative) 
statistics~\cite{Stern10rev, Ivanov01}.
A major problem for topological quantum computation is that the exchange statistics of MFs does not allow for universal 
quantum computation through braiding operations alone~\cite{Nayak08rev}. 
We showed in a recent work~\cite{Leijnse11top} that quantum information can be coherently transferred between topological qubits 
(based on MFs) and normal spin qubits. This allows using topological qubits as quantum memories in a spin-based quantum computer,
or, alternatively, to use spin-qubits to prepare arbitrary topological qubit states. 
Other works have considered interfaces between topological qubits and conventional superconducting 
qubits~\cite{Hassler10, Sau10b, Jiang11} or charge qubits defined in a double quantum dot~\cite{Bonderson11}.

Here we show that a TS can be used to directly perform entangling two-qubit gates on
spatially separated spin qubits. The two-qubit gates consist exclusively of protected braiding operations acting on a 
type of overlapping topological two-qubit states, and only the transfer of quantum information between the 
spin and topological qubits remain (partially) unprotected. 
Furthermore, unlike pulsed exchange interactions which require careful 
timing, such braiding-based two-qubit gates leave no room for continuous errors (two particles cannot be partially exchanged
and an electron cannot be partially transferred between the dots and the TS).

\section{Hybrid spin--topological qubit system}
\begin{figure}[t!]
  	\includegraphics[height=0.32\linewidth]{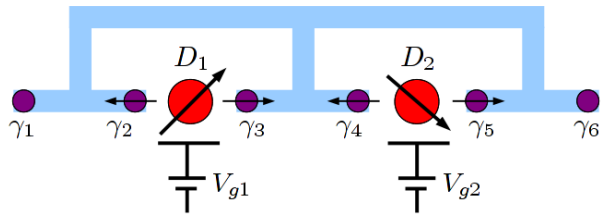}	
	\caption{(Color online) Sketch of setup. The electrostatic potentials on the quantum dots $D_1$ and $D_2$ 
	are controlled by the gate voltages $V_{g1}$ and $V_{g2}$. This allows control of their occupation, which can be either empty 
	(zero electrons on the dot) or filled (one electron on the dot), with a filled dot acting as a spin qubit. 
	The 1D topological superconducting wires host MFs described by the operators $\gamma_1, \hdots, \gamma_6$ at the
	edges. 
        The T-junctions and additional wire allow exchange (braiding) of the MFs. Arrows indicate the anti-parallel spin polarization
	of the two MFs which couple to each dot.
	As described in the main text, this setup allows two-qubit gates to be performed on the spin qubits on $D_1$ and $D_2$, which 
	can be arbitrarily far apart, as long as the MFs can be initialized, braided and measured. \label{fig:1}
 	}
\end{figure}

A sketch of the hybrid spin--topological quantum system is shown in Fig.~\ref{fig:1}. Here the TS consists of one-dimensional (1D)
wires~\cite{Oreg10, Lutchyn10}, connected to allow for particle exchange, but the results in the paper are not restricted to 
this specific geometry.
To keep the discussion as simple as possible, we focus throughout the paper on a rather simple model. The assumptions underlying this model,
as well as the effect of deviations from the model, are discussed in Sect.~\ref{sec:err}.
The Hamiltonian is then given by $H_1 + H_2$, where
\begin{eqnarray}\label{eq:HD}
	H_{i}	&=& 	\epsilon_i \sum_\sigma n_{i \sigma} + U_i n_{i \uparrow} n_{i \downarrow} \nonumber \\ 
		&+&	\gamma_{2i} \left( \lambda_{2 i} d_{i \uparrow} - \lambda_{2 i}^{*} d_{i \uparrow}^\dagger \right)  \nonumber \\ 
		&+&	\gamma_{2i+1} \left( \lambda_{2i+1} d_{i \downarrow} - \lambda_{2i + 1}^{*} d_{i \downarrow}^\dagger \right).
\end{eqnarray}

The first line describes the single-level quantum dot $D_i$ with occupation number $n_{i \sigma} = d_{i \sigma}^\dagger d_{i \sigma}$ and
onsite energy $\epsilon_i$, which can be controlled with the gate voltage, 
$\epsilon_i \propto - V_{g i}$. $U_i$ is the Coulomb charging energy, which is typically large compared with the other relevant energy scales 
and we take the limit $U_i \rightarrow \infty$. We have assumed the onsite energies to be spin-degenerate.
 
The second and third lines in Eq.~(\ref{eq:HD}) describe tunneling between the dot and the TS, see also 
e.g., Refs~\onlinecite{Leijnse11, Flensberg10b, Leijnse11top}. 
We here consider only dot energies below the superconducting gap, $|\epsilon_i| \ll \Delta$, where the tunneling is 
dominated by coupling to the zero-energy edge MFs which are described by the $\gamma$ operators. 
A MF is always spin-polarized in the sense that, for a given tunnel junction, there exist a spin-quantization axis along which 
it is only possible to tunnel into or out of the state with electrons with a specific spin projection. 
In Eq.~(\ref{eq:HD}) we have assumed this spin polarization 
to be anti-parallel for the two MFs which couple to a given dot, i.e, for $\gamma_{2 i}$ and $\gamma_{2i +1}$ 
(this assumption is motivated in Sect.~\ref{sec:err}).
We choose the quantization axis for the dot spin such that the polarization direction associated with the 
left (right) tunnel junction is spin up (spin down). This is why in Eq.~(\ref{eq:HD}), $\gamma_{2i}$ ($\gamma_{2i + 1}$) 
only couples to the spin-up (spin-down) dot electron operators. 

The MF operators are hermitian, $\gamma_i^\dagger = \gamma_i$, 
and we assume them to be normalized, $\gamma_i^2 = 1$. Thus, $\gamma_i^\dagger \gamma_i \equiv 1$ and it is not possible 
to construct a MF number operator, or to count the occupation of a Majorana mode. However, two Majoranas can be combined 
to form one ordinary fermion and we choose to pair neighboring MFs, such that 
\begin{eqnarray}
\label{gamma1}
	\gamma_{2i-1} &=& f_{i} + f_{i}^\dagger, \\
\label{gamma2}
	\gamma_{2i} &=& i(f_{i}^\dagger - f_{i}),
\end{eqnarray}
where $f_{i}^\dagger$ creates a fermion and $f_{i}^\dagger f_{i} = n_{i} = 0,1$ counts the occupation of the corresponding state.
Thus, two MFs together form a single fermionic state which can be either empty or 
filled and the 6 MFs in our system span a $2^3$-dimensional space. 
We assume vanishing coupling between the different MFs, which is satisfied when they are well separated on 
a length scale set by the effective superconducting coherence length in the TS. 
Then the MFs do not appear in the Hamiltonian, except for in the tunnel term. Because of the anti-parallel
spin polarization, there is also no effective coupling between $\gamma_{2i}$ and $\gamma_{2i + 1}$ via the quantum dots, 
and all the 8 fermionic states are therefore degenerate. 

Below we discuss in detail the different quantum computational operations which can be performed with the system in 
Fig.~\ref{fig:1}, but we want to already now give an intuitive explanation of the central aspect, namely the capability
to transfer quantum information from the spin of an electron in one of the dots to topological information encoded in 
the MFs. Let us restrict our attention to $D_1$ and the MFs 1--4 and assume that the level of $D_1$ is initially far below the 
chemical potential of the superconductor, such that the dot is occupied by a single electron (spin qubit). If we now adjust the 
gate voltage to bring up the dot level, it will eventually empty and the electron will tunnel into the TS (many electrons may 
tunnel back and forth, but there will be a net transfer of one electron from the dot to the TS). 
If the dot level is kept well inside the superconducting gap, this can only happen via tunneling into the MFs 2 and 3. 
Because of the anti-parallel spin polarizations of these MFs, the electron must split into a spin up and a spin down component 
(along the quantization axis of the MFs polarization), each component tunneling into opposite sides of the TS. 
Adding a charge to the TS to the left (right) side of $D_1$ corresponds to flipping the fermionic occupation number $n_1$ ($n_2$).
Thus, by adding a fractional charge to each side, where the fraction depends on the spin projection of the electron, we rotate 
each of the fermionic number states by an amount which depends on the electron spin. This allows translation from spin quantum 
information into topological quantum information. In the same way, we can transfer topological quantum information into 
spin quantum information.

\section{Basic computational and transfer operations}
The computational operations discussed in this paper rely on the ability to carry out a number of different operations 
on the quantum dots and MFs. We list these operations here and discuss them in more detail below.

\begin{enumerate}

\item $P_F^i$: Fill (an initially empty) dot $i$ with an electron from the TS by a gate sweep which 
shifts $\epsilon_i$ from negative to positive.

\item $P_E^i$: Empty (an initially occupied) dot $i$, transferring the electron to the TS, by a gate sweep 
which shifts $\epsilon_i$ from positive to negative.

\item $B_i$: Braid (exchange) MFs $i$ and $i+1$.

\item $I_{i}$: Initialize the fermion number states $n_{i}$ (the ability to produce $n_{i} = 0$ is sufficient).

\item $M_{i}$: Measure the fermionic occupation numbers $n_{i}$.

\end{enumerate}

1. It was shown in~\cite{Leijnse11top} that if dot $i$ is initially empty, a gate sweep which takes 
$\epsilon_i$ from $\epsilon_i / | {\bar \lambda_i} | \ll - 1$ 
to $\epsilon_i /  |{ \bar \lambda_i }| \gg 1$ has the same effect (up to an overall phase) as acting on the state of 
the total system with the operator
\begin{eqnarray} \label{eq:PF}
	P_F^i	&=&	\frac{1}{\sqrt{2}} \left( \gamma_{2i} d^\dagger_{i \uparrow} + \gamma_{2i + 1} d^\dagger_{i \downarrow} \right),
\end{eqnarray}
where we have assumed ${ \bar \lambda}_i = \lambda_{2 i} = \lambda_{2i + 1} $ (a relative phase between $\lambda_{2 i}$ and 
$\lambda_{2i + 1}$, introducing a phase in one of the terms in Eq.~(\ref{eq:PF}), does not significantly alter the results below, 
but we exclude it here for simplicity). The requirement for Eq.~(\ref{eq:PF}) to be valid is that the gate sweep is done 
adiabatically with respect to the tunnel coupling ${\bar \lambda_i}$, so that the system remains in the ground state as the dot 
is emptied.  

2. Similar to above, if dot $i$ is initially filled with an electron, an adiabatic gate sweep which takes 
$\epsilon_i$ from $\epsilon_i / | {\bar \lambda_i} | \gg 1$ 
to $\epsilon_i /  |{ \bar \lambda_i }| \ll - 1$ has the same effect (up to an overall phase) as acting with the operator
\begin{eqnarray} \label{eq:PE}
	P_E^i	&=&	N \left( \gamma_{2i} d_{i \uparrow} + \gamma_{2i + 1} d_{i \downarrow} \right).
\end{eqnarray}
There is one complication here: Eq.~(\ref{eq:PE}) assumes that the dot is successfully emptied, which is, in fact, not 
guaranteed even if the gate sweep is done adiabatically. 
The reason is that, due to the spin degeneracy, the dimension of the Hilbert space with an empty dot is smaller than that 
with a full dot. Therefore, the ground state manifold with a full dot must have a component which is disconnected from the smaller
ground state manifold with an empty dot. 
Equation~(\ref{eq:PE}) should be interpreted as the operation of making the adiabatic gate sweep, followed by a 
measurement of the charge on the dot, assuming that this charge measurement shows the dot to be empty.
However, this is not a truly restricting assumption: If the charge measurement shows that emptying the dot was not successful, 
the original spin qubit can be recovered and the transfer attempted until successful, 
see Ref.~\onlinecite{Leijnse11top} for details.
Note that, because of the collapse of the wave function, resulting from this charge measurement, the normalization $N$ 
is an operator (i.e., the correct normalization factor depends on the state upon which $P_E$ is acting).

3. The effect of physically exchanging MFs $i$ and $i+1$ is described by the operator~\cite{Ivanov01}
\begin{eqnarray} \label{eq:Bm}
	B_i	&=&	\frac{1}{\sqrt{2}}\left( 1 + \gamma_i \gamma_{i+1} \right).
\end{eqnarray}
It was recently discussed that MFs can be transported and exchanged in networks of one-dimensional 
wires using "keyboard gates"~\cite{Alicea10b}.
There are also alternatives which accomplish the same thing as braiding operations, without actual physical exchange of 
MFs~\cite{Heck12, Halperin12}. 

4. If a section of the wire is tuned from a non-topological to a topological phase, e.g., using gate voltages, one MF will 
appear in each end of the created topological region. Since such an operation does not change the parity of the fermion number 
in the superconductor, the associated fermion state must be empty, $n_{i} = 0$, see e.g.,~\cite{Alicea10b}.

5. If the MFs $2i-1$ and $2i$ are brought into proximity with each other, introducing a coupling $\xi$ between them, this 
gives rise to a term $i \xi \gamma_{2i-1} \gamma_{2i} / 2 = \xi n_{i}$ in the Hamiltonian, breaking the degeneracy associated 
with occupying the fermionic state which can then be measured~\cite{Kitaev03}. 
The occupation numbers could also be measured e.g., using interferometry~\cite{Bishara09} or by coupling to conventional 
superconducting qubits~\cite{Hassler10}.

\section{Two-qubit gates with spatially separated spin qubits}
In a slightly simpler setup than the one sketched in Fig.~\ref{fig:1}, we showed in Ref.~\onlinecite{Leijnse11top} 
that the operations described above can be used to:
\begin{enumerate}[i]
\item Transfer a spin qubit on $D_i$ into a topological qubit defined within the subspace of four MFs 
(using the operations $P_E^i I_i I_{i+1}$).
\item Transfer a topological qubit into a spin qubit ($M_i P_F^i I_i I_{i+1}$).
\item Transfer a spin qubit from one dot to the other ($M_1 P_F^2 P_E^1 I_2 I_1$)
\item Initialize the two dot spins in a maximally entangled Bell state ($M_1 P_F^2 P_F^1 I_2 I_1$).
\end{enumerate}
This can be accomplished in a setup with only fours MFs and does not require the ability to braid MFs.
We now show that, with a six-MF setup as in Fig.~\ref{fig:1}, and with the capability to braid the MFs, one can in addition
perform non-trivial \emph{two-qubit gates} on spin qubits on dots 1 and 2.

The combined state of the two dots is $|D_1 D_2 \rangle_D$, where $D_i= 0, \uparrow, \downarrow$, corresponding to dot $i$ being empty 
or filled with a spin-up or spin-down electron. We characterize the Majorana system in terms of the number states 
$|n_{1} n_{2} n_{3}\rangle_M$. 
Since braiding operations cannot change the parity of the total fermion number, four MFs are needed to form one 
topological qubit~\cite{Bravyi06}, for example within the subspace with even total fermion number parity.
It is then natural to use eight MFs to make a topological two-qubit system. However, it turns out that it is
not possible to entangle two such topological qubits using only braiding operations~\cite{Bravyi06}. 
Therefore, we cannot simply use the results of Ref.~\onlinecite{Leijnse11top} to transfer two spin qubits to two 
separate topological qubits, entangle these and then transfer the information back into the spin system.

Instead, we follow an idea by Georgiev~\cite{Georgiev06} and note that six MFs with fixed total parity
is sufficient to define a type of effective two-qubit system. If we again pick the subspace with even total 
parity we can define the two-qubit states as:
\begin{eqnarray} 
\label{eq:2qubits00}
	|{\bar 0} {\bar 0} \rangle_Q	&=&	| 0 0 0\rangle_M, \\
\label{eq:2qubits01}
	|{\bar 0} {\bar 1} \rangle_Q	&=&	| 0 1 1\rangle_M, \\
\label{eq:2qubits10}
	|{\bar 1} {\bar 0} \rangle_Q	&=&	| 1 1 0\rangle_M, \\
\label{eq:2qubits11}
	|{\bar 1} {\bar 1} \rangle_Q	&=&	| 1 0 1\rangle_M.
\end{eqnarray}
Thus, the individual qubit states can be read off from $n_1$ and $n_3$, while the fermion occupation $n_2$ is "shared" between the 
two qubits and used to balance the fermion number parity. 

In the basis 
$\{ |{\bar 0} {\bar 0} \rangle_Q, |{\bar 0} {\bar 1} \rangle_Q, |{\bar 1} {\bar 0} \rangle_Q, |{\bar 1} {\bar 1} \rangle_Q\}$, 
a straightforward evaluation of the matrix elements of Eq.~(\ref{eq:Bm}) shows that, up to overall phases, the braid 
matrices are given by
\begin{align} \label{eq:braidmatrices}
	B_1	&=	\left( \begin{array}{cccc}  i & 0 & 0 & 0 \\  0 & i & 0 & 0 \\ 
								      0 & 0 & 1 & 0 \\ 0 & 0 & 0 & 1
				  \end{array} \right), 				  
	B_2	=	\frac{1}{\sqrt{2}} \left( \begin{array}{cccc}  1 & 0 & i & 0 \\  0 & 1 & 0 & i \\
										    i & 0 & 1 & 0 \\ 0 & i & 0 & 1 
				\end{array} \right), \nonumber \\
	B_3	&=	\left( \begin{array}{cccc}  i & 0 & 0 & 0 \\  0 & 1 & 0 & 0 \\ 0 & 0 & 1 & 0 \\ 0 & 0 & 0 & i
				  \end{array} \right), 
	B_4	=	\frac{1}{\sqrt{2}} \left( \begin{array}{cccc}  1 & i & 0 & 0 \\  i & 1 & 0 & 0 \\
										    0 & 0 & 1 & -i \\ 0 & 0 & -i & 1 
				\end{array} \right), \nonumber \\
	B_5	&=	\left( \begin{array}{cccc}  i & 0 & 0 & 0 \\  0 & 1 & 0 & 0 \\ 0 & 0 & i & 0 \\ 0 & 0 & 0 & 1
				  \end{array} \right). 
\end{align}

To demonstrate how to perform two-qubit gates on the spin qubits on $D_1$ and $D_2$, we start from an initial state, 
$|i \rangle$, with one spin qubit in each dot and each fermionic number state of the TS initialized to zero:
\begin{eqnarray} \label{eq:2binitial1}
	|i \rangle = \left( \alpha_1 |\uparrow \rangle_{D_1} + \beta_1 |\downarrow \rangle_{D_1} \right) 
		     \left( \alpha_2 |\uparrow \rangle_{D_2} + \beta_2 |\downarrow \rangle_{D_2} \right) |\bar{0} \bar{0}\rangle_Q \nonumber, \\
\end{eqnarray}
where we used the rather obvious notation $|D_1\rangle_{D_1} |D_2 \rangle_{D_2} = |D_1 D_2\rangle_D$.
The first step is to empty both dots via adiabatic gate sweeps of $V_{g1}$ and $V_{g2}$, which gives the state
\begin{eqnarray} \label{eq:2bempty}
	P_E^1 P_E^2 |i\rangle	&=&	|0 0 \rangle_D \left( -\alpha_1 \alpha_2 |\bar{1} \bar{0}\rangle_{Q} + 
					i \alpha_1 \beta_2 |\bar{1} \bar{1} \rangle_{Q} \right. \nonumber \\ 
				&+&	\left. i \beta_1 \alpha_2 |\bar{0} \bar{0} \rangle_{Q} +
					\beta_1 \beta_2 |\bar{0} \bar{1} \rangle_{Q} \right) \\
 \label{eq:2bemptyfact}
				&=&
					|0 0 \rangle_D \left( \alpha_1 |\bar{1} \rangle_{Q_1} - 
								i \beta_1 |\bar{0} \rangle_{Q_1} \right) \nonumber \\
				&\times&\left( -\alpha_2 |\bar{0}\rangle_{Q_2} + i \beta_2 |\bar{1}\rangle_{Q_2} \right). 
\end{eqnarray}
Thus we see that emptying the dots has the effect of transferring the spin qubits into topological qubits:
Up to phase factors $|\uparrow \rangle_{D_1} \rightarrow |\bar{1} \rangle_{Q_1}$,
$|\downarrow \rangle_{D_1} \rightarrow |\bar{0} \rangle_{Q_1}$, and 
$|\uparrow \rangle_{D_2} \rightarrow |\bar{0} \rangle_{Q_2}$, 
$|\downarrow \rangle_{D_2} \rightarrow |\bar{1} \rangle_{Q_2}$.
Note, however, that interpreting the factorized form of the topological two-qubit states 
($|\bar{Q}_1 \rangle_{Q_1} |\bar{Q}_2 \rangle_{Q_2} = |\bar{Q}_1 \bar{Q}_2 \rangle_{Q}$) in the second equality requires some care.
The occupation number $n_2$, which is suppressed in this notation, depends on both $n_1$ and $n_3$ due to the constraint of even total parity.

If we now fill the two dots again (applying $P_F^1 P_F^2$), an entangled four-qubit state results, involving the two spin qubits 
and the two topological qubits. 
To keep the formulas short, we present the result for the case where both dots were initially initialized in the spin up state, i.e.,
$|i'\rangle = |\uparrow \uparrow\rangle_D |\bar{0}\bar{0}\rangle_Q$. 
From Eq.~(\ref{eq:2bempty}) we see that emptying both dots gives
\begin{eqnarray} \label{eq:simpleempty}
	|j\rangle &=& P_E^1 P_E^2 |i'\rangle = -|0 0 \rangle_D |\bar{1} \bar{0}\rangle_{Q}. 
\end{eqnarray}
This result is easy to understand: Since both spin-up electrons must tunnel to the left, they change the fermion occupation numbers 
$n_1$ and $n_2$ from 0 to 1. Filling both dots then gives
\begin{eqnarray} \label{eq:2bfill}
	P_F^1 P_F^2 |j\rangle	&=& -\frac{1}{2} \left( |\uparrow \uparrow\rangle_D |\bar{0}\bar{0}\rangle_Q 
				    + |\downarrow \downarrow\rangle_D |\bar{1}\bar{1}\rangle_Q \right. \nonumber \\
				&-&  \left. i |\downarrow \uparrow\rangle_D |\bar{1}\bar{0}\rangle_Q
				    -i |\uparrow \downarrow\rangle_D |\bar{0}\bar{1}\rangle_Q \right).
\end{eqnarray}
We can get back the spin qubits by collapsing the wavefunction of the topological qubits, which 
is accomplished by measuring the topological qubits $Q_1$ and $Q_2$ (by measuring $n_1$ and $n_3$). 
This gives four possible outcomes with equal probability, each of which describes two spin qubits. As is seen from Eq.~(\ref{eq:2bfill}),
depending on the measured values of $n_1$ and $n_3$, one or both spin qubits may have been flipped compared with the initial state
(note that compared with $|i'\rangle$, a flip of the spin on dot 1 (2) is always accompanied by a change of $n_1$ ($n_3$)).
However, the spin qubits are still \emph{always in a product state}, also for a general initial state like $|i\rangle$.
This result demonstrates that two spin qubits can be simultaneously saved in the overlapping topological two-qubit system and then 
recovered. However, it also shows that transfer operations alone are insufficient for performing entangling two-qubit gates on the spin qubits. 

What is needed is to entangle the topological qubits $Q_1$ and $Q_2$, before transferring them back 
into the spin system. Fortunately, braiding operations can create entangled states in the overlapping 
two-qubit basis~(\ref{eq:2qubits00})--(\ref{eq:2qubits11}). From Eq.~(\ref{eq:braidmatrices}) it is clear that the operations 
$B_2$ and $B_4$ have the effect of creating an entangled state when acting on a two-qubit product state. These are the 
braiding operations which involve one
of the shared MFs (which together form $n_2$) with one of the end MFs making up $n_1$ or $n_3$. In contrast, the other braiding operations, 
which involve exchange of MFs within each qubit, merely result in single-qubit rotations. 

Thus, a non-trivial (entangling) two-qubit gate is e.g., obtained by the combined operations 
$M_1 M_3 P_F^1 P_F^2 B_2 P_E^1 P_E^2 I_1 I_2 I_3$ 
(initialize Majorana number states, empty dots, braid MFs 2 and 3, fill dots, measure $n_1$ and $n_3$). 
By combinations of braiding operations we can construct e.g., a CNOT gate~\cite{Georgiev06} acting on the topological 
two-qubit system. Applying the CNOT gate to the topological part of the state $|j\rangle$ in Eq.~(\ref{eq:simpleempty}) 
before filling the dots gives
\begin{eqnarray} \label{eq:CNOTfill}
	P_F^1 P_F^2 \; \text{CNOT}\; |j\rangle &=& - P_F^1 P_F^2 |00\rangle_D |\bar{1} \bar{1}\rangle_Q \nonumber \\
					       &=& -\frac{1}{2} \left( |\uparrow \uparrow\rangle_D |\bar{0}\bar{1}\rangle_Q 
				    		   - |\downarrow \downarrow\rangle_D |\bar{1}\bar{0}\rangle_Q \right. \nonumber \\
					       &+&  \left. i |\downarrow \uparrow\rangle_D |\bar{1}\bar{1}\rangle_Q
				    		   -i |\uparrow \downarrow\rangle_D |\bar{0}\bar{0}\rangle_Q \right).
\end{eqnarray}
After measuring $n_1$ and $n_3$ we are of course still left with four possible outcomes; if $n_1 = n_3 = 0$ is measured
a CNOT gate has been performed on the spin qubits, while other measured $n_1$ and $n_3$ correspond to other gates. 
However, as was briefly mentioned above, these other gates are related to the $|\bar{0}\bar{0}\rangle_Q$-gate 
by single spin-qubit rotations. It holds in general that to retrieve the result of the $|\bar{0}\bar{0}\rangle_Q$-gate (up to a phase factor), 
one should perform a $\pi$-rotation around the $y$-axis of the spin on dot $1$ if $n_1 = 1$ and on dot 2 if $n_3 =1$. Thus, in combination with 
the parity measurements, (possibly) followed by single-qubit rotations, our proposed two-qubit spin gates are fully deterministic.
The result in Eq.~(\ref{eq:CNOTfill}) may seem trivial, but
starting from the general two-qubit state $|i\rangle$ rather than $|i'\rangle$, one sees that this procedure indeed
produces a CNOT gate, up to single-qubit rotations.

\section{Physical assumptions and sources of errors\label{sec:err}}
Finally, we discuss the assumptions made when writing down the model in Eq.~(\ref{eq:HD}), as well as the effect on the results
when the most crucial assumptions are not perfectly fulfilled.  

First, we have included only a single level on each dot and in addition let $U_i \rightarrow \infty$. 
The likely most experimentally attractive option is to construct the quantum dots from the same 1D wires as the TS,
but not covered by a superconductor and separated from the TS for example by electric gates. The dots would then 
be defined in e.g., InAs or InSb nanowires (good candidates for 1D TS). Such dots have been found to exhibit level spacings 
and Coulomb charging energies of several meV, as discussed e.g., in Refs.~\onlinecite{Fasth07, Nilsson09}.
Since this is much larger than both the expected TS gap ($\lesssim 1$ meV) and operating temperature ($\lesssim 100$~mK),
these assumptions are rather unproblematic.

Second, the dots are assumed to be spin degenerate. This could be achieved if
the Zeeman term needed to induce topological superconductivity is provided by proximity with a magnetic insulator, 
see e.g., Refs.~\onlinecite{Fu08, Sau10}, rather than induced by an external magnetic field. 
Alternatively, a magnetic field could be tuned to recover spin degeneracy, but between spin-up and spin-down states 
in different dot orbitals. 
It has also been suggested~\cite{Stoudenmire11} that strong interactions may lower the magnetic field needed to produce 
Majorana fermions, or even cause them to appear at zero field due to spontaneous breaking of time-reversal symmetry.
Finally, the $g$-factor on the dot can be much smaller than that in the TS (even if they are made from the same material since 
strong confinement often renormalizes the $g$-factor), making it possible to reach the topological regime while introducing 
only a small Zeeman splitting on the dot. In fact, a small Zeeman splitting, $\delta B \ll \lambda$, is acceptable. A gate sweep 
which is adiabatic with respect to $\lambda$, but non-adiabatic with respect to $\delta B$, then still results in the same filling and 
emptying operators, Eqs.~(\ref{eq:PF}) and~(\ref{eq:PE}), see Ref.~\onlinecite{Leijnse11top}.

Finally, we have assumed perfect anti-parallel spin polarizations of the two MFs which couple to a given dot. 
Under idealized conditions, for some realizations of MFs, the spin polarization direction is controlled 
by the directionality of the border between topological and non-topological regions and would therefore naturally be 
opposite for opposite edges as sketched in Fig.~\ref{fig:1}, see e.g., Refs.~\onlinecite{Oreg10, Fu09,Shivamoggi10, Fu08}.
Under non-ideal conditions, the polarization direction will depend on the details of the edge,
as discussed in some detail e.g., in Refs.~\onlinecite{Kjaergaard12, Sticlet12}.
Tuning the MF spin polarization can e.g., be achieved through control of the chemical potential close to the edge, 
which must likely be experimentally controllable to reach the topological regime anyway. 
If the dot level is tuned close to resonance, a deviation from anti-parallel polarizations lead to a finite supercurrent through 
the system, which can therefore be used as an indicator when fine tuning to the anti-parallel configuration.

If perfect anti-parallel spin polarizations cannot be achieved, all is not lost. The supercurrent and the associated splitting 
of the TS ground state degeneracy can be eliminated by control of the relative phase difference between the TS on the two sides of the 
dot~\cite{Flensberg10b}. Filling and emptying dot $i$ is then described by the operators [cf. Eqs.~(\ref{eq:PF}) and~(\ref{eq:PE})]
\begin{eqnarray} 
\label{eq:PFimperfect}
	\tilde{P}_F^i	&=&	\frac{1}{\sqrt{2}} \left[ \gamma_{2i} \left( d^\dagger_{i \uparrow} \; \mathrm{cos} \frac{\theta}{4} 
				+ d^\dagger_{i \downarrow} \; \mathrm{sin} \frac{\theta}{4}\right) \right. \nonumber \\
			&+&  	\phantom{\frac{1}{\sqrt{2}}}\left. 
				\gamma_{2i + 1} \left( d^\dagger_{i \downarrow} \; \mathrm{cos}\frac{\theta}{4} 
				+ d^\dagger_{i \uparrow} \; \mathrm{sin}\frac{\theta}{4} \right) \right], \\
\label{eq:PEimperfect}
	\tilde{P}_E^i	&=&	N \left[ \gamma_{2i} \left( d_{i \uparrow} \; \mathrm{cos} \frac{\theta}{4} 
				+ d_{i \downarrow} \; \mathrm{sin} \frac{\theta}{4}\right) \right. \nonumber \\
			&+&  	\phantom{\frac{1}{\sqrt{2}}}\left. 
				\gamma_{2i + 1} \left( d_{i \downarrow} \; \mathrm{cos}\frac{\theta}{4} 
				+ d_{i \uparrow} \; \mathrm{sin}\frac{\theta}{4} \right) \right], 
\end{eqnarray}
where $\theta$ is the angular deviation from perfect anti-parallel spin polarizations. Acting with $\tilde{P}_F^i$ or $\tilde{P}_E^i$ still
accomplishes a quantum information transfer between topological qubits and spin qubits, but introduces an error 
$\delta = \mathrm{sin} \frac{\theta}{4}$ (compare with the ideal cases described by Eqs.~(5) and~(9) in Ref.~\onlinecite{Leijnse11top}).
The two-qubit gates discussed above require only two empty and two fill operations. At least for a proof of principle experiment, a rather sizable
spin mis-alignment is therefore acceptable.

\section{Conclusions}
The setup in Fig.~\ref{fig:1} could easily be extended to include an arbitrary number of quantum dots coupled to edge Majorana 
bound states. 
In such a geometry, the combined results described here and in Ref.~\onlinecite{Leijnse11top} allow transferring spin qubits between arbitrary 
dots, initializing any given pair of spin qubits in a maximally entangled Bell state, and performing non-trivial 
two-qubit gates on arbitrary spin qubits. Furthermore, the interface with the topological qubits provides a quantum memory 
with potentially very long coherence times. Together with the fast manipulation of single spin qubits or spatially close 
qubit pairs, known from conventional spin qubit systems, this provides an interesting platform for quantum computation with spins.

In addition to the possibility of acting on spatially separated spin qubits, our proposed topologically-assisted two-qubit gate has 
the advantage of avoiding any continuous errors. The braiding-, transfer-, and measurement operations can be done to yield 
exactly the desired operation, unlike a controlled exchange interaction which requires very accurate timing and unavoidably 
introduces small errors.
While the braiding operations are being performed, both dots are empty and all information is stored in the topological system 
and therefore benefits from the associated (hopefully) long coherence times. Only the transfer operations are unprotected, but, 
assuming that the braiding operations can be carried out without decoherence, we have reduced the problem of performing a 
two-spin-qubit gate to that of gate-controlled transfer of two individual charges. 

Note that the problem of generalizing such an overlapping qubit system to more than two qubits~\cite{Georgiev06, Bravyi06}
is not an issue in our case, since at any given time, at the most two spin qubits have to be stored in the topological system, and
any other spin-qubits (on other dots) are of course unaffected by the braiding and transfer operations done to these qubits.

\bibliographystyle{apsrev}
\end{document}